\documentclass[conference]{IEEEtran}
\IEEEoverridecommandlockouts
\usepackage{cite}
\usepackage{amsmath,amssymb,amsfonts}
\usepackage{algorithmic}
\usepackage{graphicx}
\usepackage{textcomp}
\usepackage{xcolor}
\def\BibTeX{{\rm B\kern-.05em{\sc i\kern-.025em b}\kern-.08em
    T\kern-.1667em\lower.7ex\hbox{E}\kern-.125emX}}
\begin{document}

\title{The Ivory Tower Lost: How College Students Respond Differently than the General Public to the COVID-19 Pandemic
}


\author{\IEEEauthorblockN{Viet Duong, Jiebo Luo}
\IEEEauthorblockA{\textit{Department of Computer Science} \\
\textit{University of Rochester}\\
Rochester, USA \\
vduong@ur.rochester.edu, jluo@cs.rochester.edu}
\and
\IEEEauthorblockN{Phu Pham, Tongyu Yang}
\IEEEauthorblockA{\textit{Goergen Institute for Data Science} \\
\textit{University of Rochester}\\
Rochester, USA \\
\{ppham2, tyang20\}@u.rochester.edu}
\and
\IEEEauthorblockN{Yu Wang}
\IEEEauthorblockA{\textit{Political Science }\\
\textit{University of Rochester}\\
Rochester, USA \\
ywang176@ur.rochester.edu}
}


\maketitle

\begin{abstract}
Recently, the pandemic of the novel Coronavirus Disease-2019 (COVID-19) has presented governments with ultimate challenges. In the United States, the country with the highest confirmed COVID-19 infection cases, a nationwide social distancing protocol has been implemented by the President. For the first time in a hundred years since the 1918 flu pandemic, the US population is mandated to stay in their households and avoid public contact. As a result, the majority of public venues and services have ceased their operations. Following the closure of the University of Washington on March 7th, more than a thousand  colleges and universities in the United States have cancelled in-person classes and campus activities, impacting millions of students. This paper aims to discover the social implications of this unprecedented disruption in our interactive society regarding both the general public and higher education populations by mining people's opinions on social media. We discover several topics embedded in a large number of COVID-19 tweets that represent the most central issues related to the pandemic, which are of great concerns for both college students and the general public. Moreover, we find significant differences between these two groups of Twitter users with respect to the sentiments they expressed towards the COVID-19 issues. To our best knowledge, this is the first social media-based study which focuses on the college student community's demographics and responses to prevalent social issues during a major crisis.
\end{abstract}


\section{Introduction}

First detected in Wuhan, China on December 31th, 2019, COVID-19, or the coronavirus, outbreak grew rapidly in scale and severity, and was officially declared as a pandemic on March 11th, 2020\footnote{https://edition.cnn.com/2020/02/06/health/wuhan-coronavirus-timeline-fast-facts/index.html}. As of April 13th, the World Health Organization (WHO) reported 1,812,734 confirmed cases of COVID-19 worldwide, including 113,675 deaths\footnote{https://who.sprinklr.com/}. Due to the novelty and intractability of the virus, the global community, particularly the elderly and those with underlying medical problems\footnote{https://www.who.int/health-topics/coronavirus}, are at a high risk for serious health and safety hazard. However, we suspect that the younger and physically healthier population is just as susceptible, though in a different way, to COVID-19. In order to control the spread of the outbreak, non-pharmaceutical interventions and preventive measures such as social-distancing and self-isolation have been implemented worldwide out of utmost necessity, which has led to the large-scale shutdown of public gathering places. As members of an active working and learning society, people who dedicate most of their daily hours at workplaces and educational institutions are highly vulnerable to the impacts of the closure of these facilities.

This is especially true for college students. The response to the COVID-19 pandemic has brought a sudden disruption in the operations of schools, colleges and universities, influencing more than 1.7 billion students in 192 countries\footnote{https://en.wikipedia.org/wiki/Impact\_of\_the\_2019-20\_coronavirus\_pandemic\_on\_education}. Located at the epicenter of the pandemic, with 579,005 confirmed cases, including 22,252 deaths\footnote{https://www.cdc.gov/coronavirus/2019-ncov/cases-updates/cases-in-us.html}, the U.S educational system, which is one of the largest in the world, has taken the biggest hit. Beginning with the University of Washington, which closed its campus on March 7th, 2020 and moved classes online for its 50,000 students, many colleges have also immediately responded to this outbreak by cancelling all on-campus activities such as workshops, conferences, and sports, and relocating their in-person classrooms to online platforms. As of April 14th, 2020, more than 124,000 U.S. public and private schools have closed due to the virus, affecting at least 55.1 million students\footnote{https://www.edweek.org/ew/section/multimedia/map-coronavirus-and-school-closures.html}. This transition has introduced multiple challenges for students. The foremost concern is related to how the government and education system handle the pandemic crisis with the study-from-home approach. Past surveys suggested that students experience severe limitation on particular subjects that benefit from physical interaction with the materials, and tend to lose the "pacing mechanism" of scheduled lectures, thus have a higher chance of dropping out than those in traditional settings \cite{fedynich2013teaching, morrison2019designing}. More recently, as the COVID-19 pandemic is unfolding, Sahu~\cite{sahu2020closure} hinted at other issues related to the closure of schools, apart from online learning, such as international students' travel and students' mental health. This motivates us to provide a more comprehensive study on the student demographics regarding their primary subjects of concern and how they are expressed, particularly amidst the COVID-19 crisis.

In this study, we attempt to explore the responses to the COVID-19 pandemic by Twitter users, with the focus on the college students. Also, we highlight our findings regarding the college student demographics by characterizing the outstanding differences in their behaviors from the general public. Such insights can be vital for educators and policy-makers to measure the effectiveness of their on-going efforts in the global fight against COVID-19 and the protection of our younger population. In addition, we train classification models to identify the demographics of users who posted tweets associated with COVID-19, as well as extracting the sentiments they inherently expressed in their posts. The models can be used in social media platforms to investigate central social problems, with respect to both their universality and degree of impact, and to draw the community's attention towards high-priority targets for addressing such problems.

Our main contributions are several folds: (1) we approach the social issues related to COVID-19 across different demographics using social media, by collecting data from Twitter; (2) we deploy topic modeling methods on a novel dataset to highlight the topical patterns of the ongoing social media discussions during a major crisis, regarding two different user demographics on Twitter; and (3) our implementation of state-of-the-art transformer models for natural language inference achieves new state-of-the-art performance for the Twitter sentiment classification task, which allows meaningful insights on social media behaviors to be discovered reliably.

\section{Related Work}
Our study draws knowledge from the body of research on characterizing the demographics of social media users, along the dimensions such as gender \cite{rao2010classifying, bergsma2013using}, age \cite{nguyen2013old, sloan2015tweets}, and social class \cite{sloan2015tweets}. Methods on inferring the Twitter user demographics were typically reliant on mining fine-grained linguistic patterns from the user's Twitter biography (short self-descriptive text) and posts, which are proven highly precise for certain attributes when properly constructed \cite{bergsma2013using, beller2014ma}. These approaches have also been deployed with relatively strong performance for the college student demographics \cite{hanson2013tweaking, he2016using}. Due to the recent advance in neural networks for sequence and image classification, Wang et al.~\cite{wang2019demographic} were able to leverage a multimodal, multi-attibute, and multilingual approach to achieve the  state-of-the-art accuracy on gender, age, and organization entity classification. Building upon the discoveries of previous works, we design and evaluate our own college student user classification method. This enables us to identify the two pools of users (college students and general public) among college followers on Twitter for the subsequent comparative analysis.

Research on sentiment analysis for Twitter textual data, which tackles the problem of analyzing the messages posted on Twitter in terms of the sentiments they express, has also been performed. Twitter is a very challenging domain
for sentiment analysis mainly due to the length limitation of texts \cite{giachanou2016like}. The majority of past approaches employed a traditional machine learning methods such as logistic regression, SVM, MLP, etc., trained on lexicon features and sentiment-specific word embeddings (vector representations of words) \cite{korenek2014sentiment, tang2014learning}. More recent approaches typically opted for sequence learning models trained to learn relevant embeddings for classification from large pretrained word embeddings, particularly the GloVe \cite{pennington2014glove} embeddings for Twitter data. Best performing models of this breed include Cliche (2017) \cite{cliche-2017-bb} and Baziotis et al. (2017) \cite{baziotis-etal-2017-datastories-semeval}, which shared the first place for sentiment analysis on Twitter (Task 4A \cite{rosenthal2019semeval}) at the International Workshop on Semantic Evaluation 2017 (SemEval-2017). The novelty in our approach to Twitter sentiment analysis involves the implementation of state-of-the-art transformer methods such as BERT \cite{devlin-etal-2019-bert} and RoBERTa \cite{liu2019roberta}, whose outstanding sentiment classification prowess remains untested for Twitter data in the literature.

\section{Data Collection and Preprocessing}
\subsection{Data Collection}
In this study, we limit the user population to those who follow the official Twitter accounts of colleges in the U.S. News 2020 Ranking of Top 200 National Universities. Relevant users identified as English speakers were collected using the Tweepy API\footnote{https://git.io/JvAjh}. Since the lists of the followers of the colleges in consideration might overlap, simply collecting tweets from these users could create major data redundancy and time complexity problems. Therefore, we extract unique users from the combined results, as well as their personal information and profile images, to obtain a dataset of 12,407,254 unique users. This set of users is relatively large and 1,641,582 of of these users have Twitter protected accounts, which means we are not allowed to collect tweets (Twitter posts) from them. Thus, we randomly sample 100,000 users from the unprotected pool to represent the population of college followers for the subsequent tweet collection and text analysis.

Tweets were collected using the Tweepy API. We retrieved a total of 1,873,022 tweets from the 100,000 user samples posted within the timeframe between January 20th, when the first COVID-19 case was confirmed in the U.S., and March 20th of 2020 to cover a two-month period, when nationwide social distancing protocol and school closure were attracting mass concerns. We then extracted tweets related to COVID-19, with a list of keywords consisted of "corona", "\#Corona, "\#coronavirus", "covid-19", "covid19", "coronavirus", "\#Covid\_19", "chinese virus", and "\#ChineseVirus". As a result, we obtain 73,787 unique COVID-19 related tweets, pertaining to 12,776 users, whom in this study we will address as \textit{affected users}. In addition, tweets that are not related to COVID-19 of the 12,776 \textit{affected users} are kept for the student inference task.

\subsection{Text Preprocessing} 
We develop a text preprocessing pipeline similar to that of Baziotis et al.~\cite{baziotis-etal-2017-datastories-semeval} to ensure that our text dataset is to a high degree lexically comparable to natural language, and include COVID-19 domain-specific word knowledge from a novel dataset. This is done by performing sentiment-aware tokenization, spell correction, word normalization, segmentation (for splitting hashtags) and token annotation. They implemented a tokenizer with the \textit{SentiWordnet} corpus \cite{esuli2006sentiwordnet}, which is capable of avoiding splitting expressions or words that should be kept intact (as one token), and identify most emoticons, emojis, expressions such as dates, currencies, acronyms, censored words (e.g. s**t), etc. In addition, we perform spelling correction on the extracted tokens by composing a dictionary for the most commonly seen abbreviations, censored words and elongated words (for emphasis, e.g. "reallyyy"). The \textit{Viterbi} algorithm is used for word segmentation, with word statistics (unigrams and bigrams) computed from a recently published Twitter dataset of 50 million English tweets related to COVID-19 \cite{chen2020covid}, to obtain the most probable segmentation posteriors. Moreover, all texts are lower-cased, while URLs, emails and mentioned usernames are annotated with common designated tags and removed to retain the natural language elements from the text data. The processed tweets are then annotated by the Standford CoreNLP English annotator \cite{manning-etal-2014-stanford}, which uses syntactic constituency and dependency tree parsing to extract the appropriate part-of-speech (POS) tags and lemmas (the base/dictionary forms of words) from the tweet tokens.

\section{inference of College student demographics}
\subsection{Extracting Age, Gender and Organization Attributes from Twitter User Profiles}
We consider age, gender and organization entity to be highly descriptive attributes to first obtain a general view of our user samples. According to National Center for Education Statistics (NCES) \footnote{https://nces.ed.gov/}, as of Fall 2017, 56.2\% of enrolled students aged between 19 and 29 years old, 20.1\% are under 18, and 56.6\% of them were female. These student demographic statistics are projected by NCES to remain consistent through 2020. Also, organizational Twitter accounts apparently should not be targeted for student inference because college students are individuals. These attributes are extracted using the M3 (Multilingual, Multimodal, Multi-attribute) deep learning system for inferring the demographics of users from four sources of information from Twitter profiles: user's name (first and last name in natural language), screen name (Twitter username), biography (short self-descriptive text), and profile image \cite{wang2019demographic}. We extract 1,111 organization entities from 12,776 \textit{affected} users, and disregard them from comparative analysis since they are not individuals. Also, the gender and age attributes are used to verify our classification results.

\begin{figure}[htbp]
\centerline{\includegraphics[scale = 0.3]{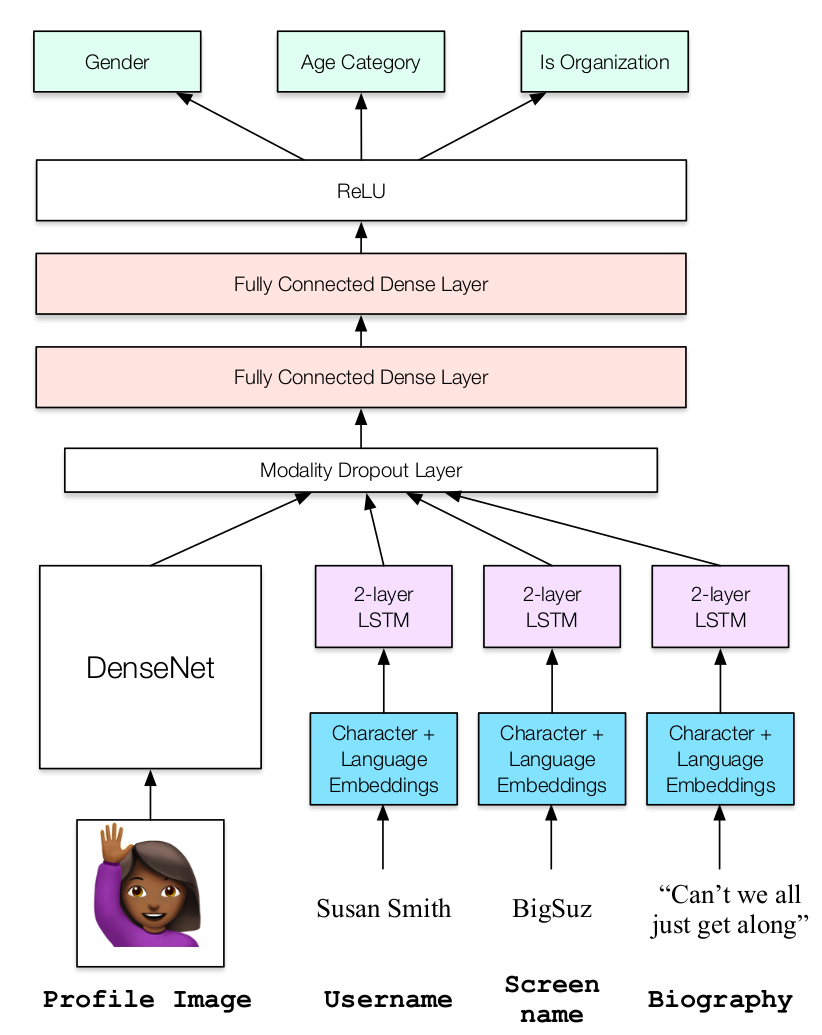}}
\caption{The M3 Model for Inferring Gender, Age, and
Organization-identity from Image and Text Data \cite{wang2019demographic}.}
\label{fig1}
\end{figure}

Although the M3 Model is highly robust for gender (0.918 Macro-F1) and organization entity (0.898 Macro-F1) recognition without using tweets, the distinguishing features for these attributes are widely available in the name, profile image, and description of Twitter users, which is not necessarily the case for the college student demographics. Since attributes from the user's name and photo image are not necessarily indicative of college students, we attempt to retrieve the students by matching the word "student" with the Twitter biographies, and found only 248 matches (2.22\% of the non-organizational users). We also matched the keywords directly related to their degree status (such as BS, MS, MBA, PhD, etc.) to their Twitter biography, and found 335 (3\%). While college students are likely to mention their degree status rather than current occupation, and many users mention the names of colleges in their biography, these should not be deciding factors because they might just be college alumni or professors instead of actual students. Due to the fact that the users are already college followers, we expect much higher percentages of college students. Thus, the use of tweets for college student identification is critical to our study.

\subsection{Heuristically Identifying College Students Using Tweets}
\subsubsection{Gold-Standard Annotations}
We sample 2,400 random users from the 11,165 non-organizational \textit{affected users} and includes their names, profile images, biographies, and tweets from 1/20 to 3/20/2020. This information is used by human annotators\footnote{Annotators are IRB certified for Social-Behavioral-Educational Research} to answer the prompt: "Would you think this person is a COLLEGE STUDENT?" with two response options: "Yes" or "No".

\subsubsection{Supervised Classification}
We encode the standard Bag of N-grams (for 1 up to 4-grams) representation of the user's tweets, which has been highly effective in text categorization and information retrieval \cite{sebastiani2002machine}, to use them as features for our classifiers. To increase the generality of our Bag of N-grams features, we preprocess the tweets as described above and apply TF-IDF vectorization, a term re-weighting scheme that discounts the influence of  common terms. We train a Random Forest classifier and report the accuracy: the percentage of correctly labeled users on 20\% of the labeled samples.

\subsubsection{Using Heuristic to Override the Classifier}
Regarding the self-distinguishing attributes of Twitter users from tweets, Bergsma and Van Durme~\cite{bergsma2013using} discovered that users most frequently reveal their attributes in the possessive construction, that is “my X” where X is an attribute, quality or event that they possess (in a linguistic sense). As a matter of fact, we found 306 tweets with the phrase ``my class" among the 1,156,947 tweets from non-organizational users. On the contrary, phrases like "I have/had (a) class(es)" occur only 16 times. Therefore, we extract this "my X" attribute type for the college student demographic as follows: we first part-of-speech tag our data using the Stanford CoreNLP tagger and then look for “my X” patterns where X is a sequence of tokens terminating in a noun. To calculate the association between the attributes and the college student demographic, we compute the pointwise mutual information \cite{church1990word} between each attribute A and student over the set of occurrences. If $PMI>0$, the observed probability of a student and attribute co-occurring is greater than the probability of co-occurrence that we would expect if student and attribute A were independently distributed.

\begin{equation} 
\label{eqn:pmi}
  PMI(A,student) = \log \frac{p(A, student)}{p(A)p(student)}
\end{equation}

We employ two techniques for selecting distinctive attributes for college students: (1) we rank the attributes by their PMI scores and use a threshold to select the top-ranked attributes; (2) we manually filter the remaining set of attributes to select those that are judged to be discriminative, including phrases closely associated with college students such as "my zoom class", "my professor", "my dorm", etc. Then we use a simple heuristic to use our identified self-distinguishing attributes in conjunction with a classifier trained on gold-standard annotations: If the user has any self-distinguishing "my-X" attributes, we assign the user to be a college student; otherwise, we trust the output of the classifier. We apply this rule to bootstrap the knowledge learned by the classifier in conjunction with our domain-specific attributes to automatically label the unannotated users. In the end, we verify the performance of our heuristic on the same test set as the classifier.

\subsubsection{Summary of Results}
As previously discussed, the Random Forest classifier, trained on 1,920 examples, performs quite well with Bag of N-grams features by correctly labeling \textbf{78\%} of the college students on the test set. We experiment with our "my-X" attributes and set the PMI threshold to 0.5, and then manually filter out the irrelevant attributes. Applying our heuristics to override the classifier improve the accuracy further to \textbf{83\%}. Therefore, we have firm grounds to utilize the combined classifier and "my-X" heuristics to label college students from the remaining users, which account for an additional 2,575 out of the total of 3,460 college student users (31\% of the non-organizational users).

\begin{figure}[htbp]
\centerline{\includegraphics[scale = 0.42]{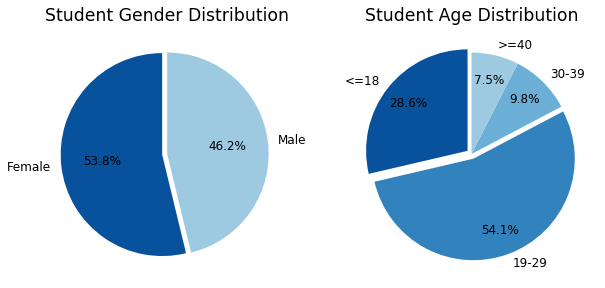}}
\caption{Gender and Age Distributions of 3,460 College Student Users.}
\label{fig2}
\end{figure}

Looking at the age and gender distributions of the college students in our samples (Figure \ref{fig2}), the statistics are very consistent with real world data in the U.S. as 53.8\% of the college students we identified are female, which is very close to the 56.7\% female percentage predicted by NCES for 2020. Although age classification is a challenging task for the M3 model (0.522 Macro-F1) and even human \cite{wang2019demographic}, our results are still within a reasonable margin with NCES's 2020 projection, with 54.1\% of the students in the 19-29 age group (vs. 56.7\%) and 28.6\%  under 18 (vs. 21.2\%).

\section{Topical Analysis of Covid-19 Tweets}

\begin{figure*}[htbp]
\centerline{\includegraphics[scale = 0.301]{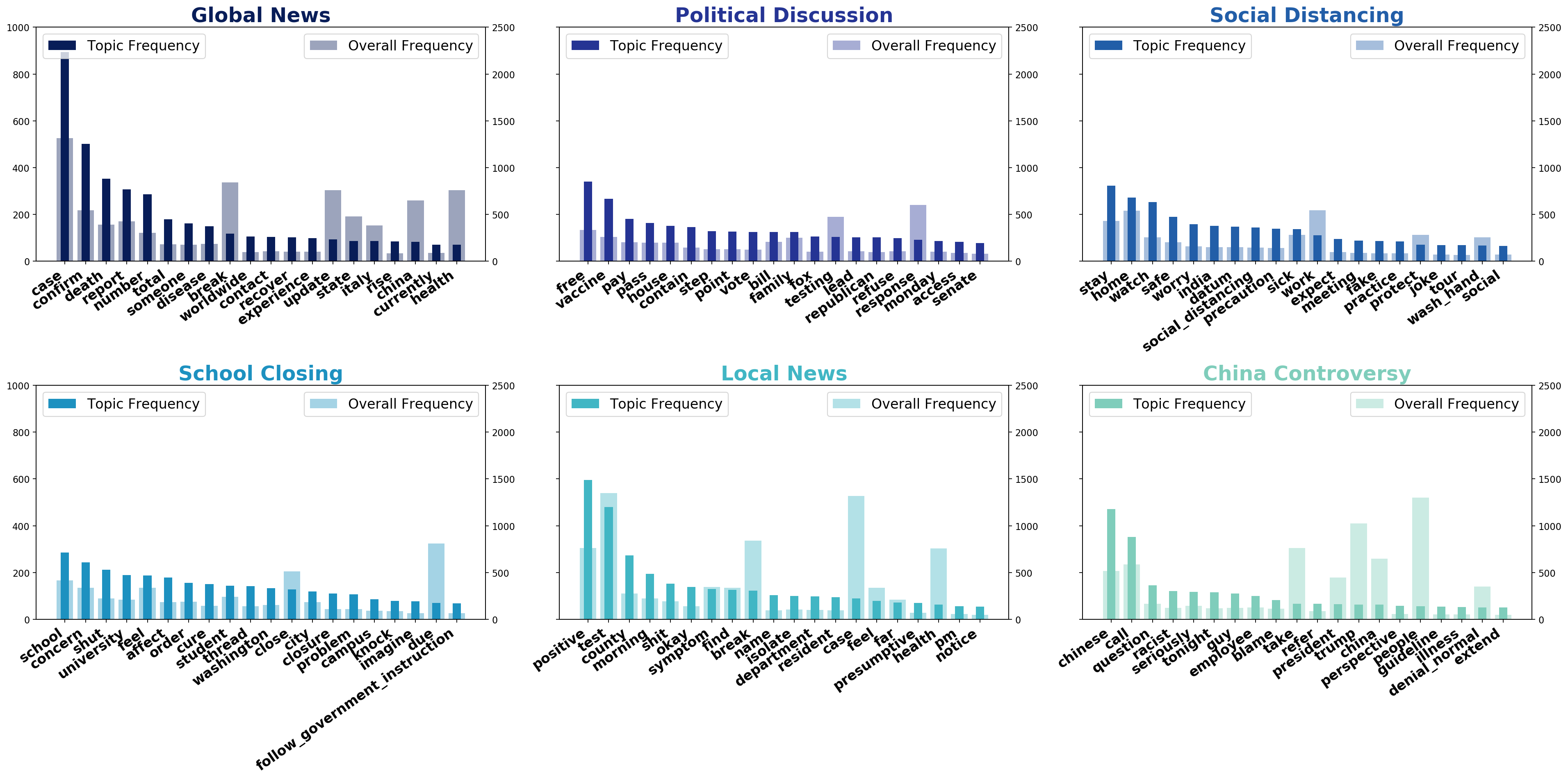}}
\caption{Topic-wise and Overall Frequency of the Top 20 Topic Keywords}
\label{fig4}
\end{figure*}

\begin{figure}[htbp]
\centerline{\includegraphics[scale = 0.28]{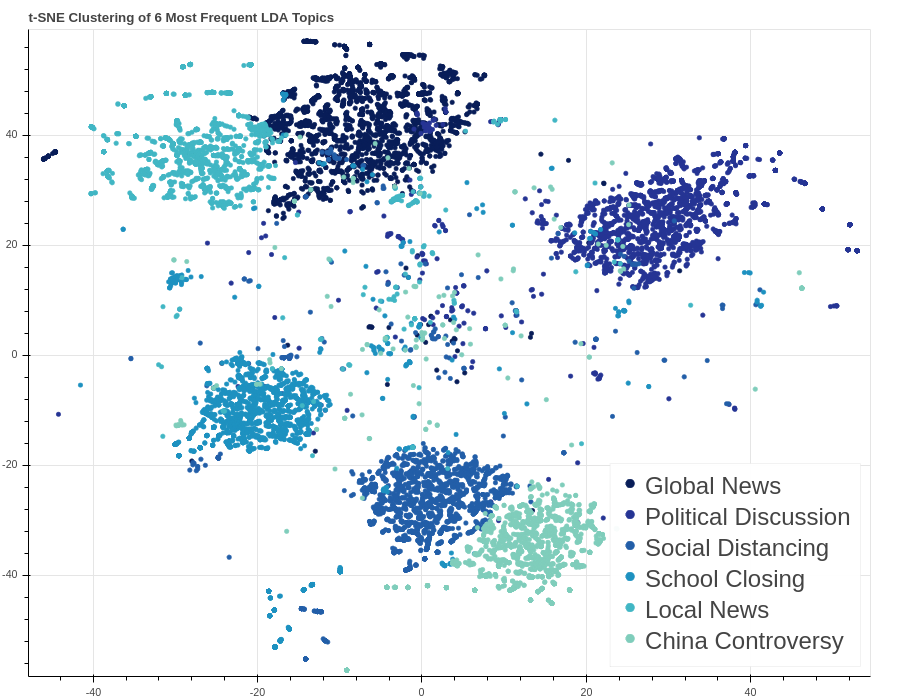}}
\caption{t-SNE Clustering of 6 Most Frequently Discussed LDA Topics.}
\label{fig3}
\end{figure}

In order to understand the latent topics of the COVID-19 tweets for college followers, we utilize Latent Dirichlet Allocation (LDA) \cite{NIPS2010_3902} to label universal topics demonstrated by the users. To reduce the complexity of the LDA analysis corpus, only the lemmas of tokens with POS tags of type noun, verb, adjective, and adverb are kept from the preprocessed tweets, because they possess the most meaningful contents related to the topics we are looking to discover. We not only look at individual tokens but also consider highly-correlated groups of two and three words. Thus, bigrams and trigrams are computed and added to the corpus. Since certain terms frequently appear in those COVID-19 tweets (e.g. virus, disease, infection, case, test, etc.), we transform our LDA corpus using TF-IDF vectorization. We finetune our LDA topic model and arrive at the optimal topic number of 55 and coherence score of 0.373. We also implement t-SNE dimensionality reduction technique \cite{maaten2008visualizing} to transform the computed 3126-dimensional document-term topic posterior matrix into 2-dimensional data points. This allows us to observe a distinctive separation of the data points representing the topic clusters, as illustrated in the plot of the 6 most prevalent topics of the COVID-19 tweets (Figure \ref{fig3}).

We then label the 6 most frequently discussed topics using the top 20 weighted topic keywords (Figure \ref{fig4}). Evidently, global news is the most popular topic among the tweets, as the numbers of confirmed positive COVID-19 cases and deaths are constantly increasing globally. The presence of political discussions, as well as the controversy related to the Chinese origin of the virus, is very strong, due to the ongoing presidential election campaign in the US, which gives solid evidence that the COVID-19 pandemic is influencing our political picture. The third and fourth most frequent topics involve social distancing and the closure of colleges.

\begin{figure}[htbp]
\centerline{\includegraphics[scale = 0.33]{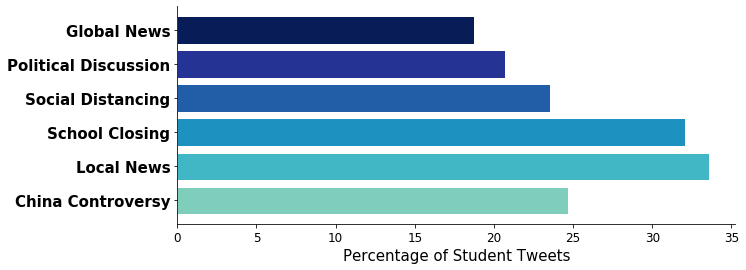}}
\caption{Student Tweets Contribution towards the Top 6 Topics.}
\label{fig5}
\end{figure}

Regarding which of the topics attracts the most attention from the college community in comparison with the general public, we find that college students tend to give more responses to COVID-19 issues that particularly affect them. In other words, as most universities in the U.S announced the shutdown of on-campus activities and encouraged students to refrain from crowded travel and commute during March, colleges students posted more tweets related to school closure (32.04\%) and news from areas close to their living proximity (33.56\%). In addition, they were concerned about social distancing and controversies regarding the address of the COVID-19 virus as "Chinese virus", which are two very important topics that we will take a closer look in later analysis.

\section{Topic-based sentiment analysis}
To expand the scope of our study from the topic modeling results, we decide to dive deeper into the posts belonging to the each of the 6 most frequently discussed topics. Specifically, for each topic, we separate the college students and general population into two pools and apply the RoBERTa model to classify and examine the sentiments they expressed. Also, we use the same topic modeling techniques as described in Section V to provide microscopic explanations to the sentiment results.

\subsection{ Transformer Models for Sentiment Classification}
\subsubsection{RoBERTa - Robustly Optimized BERT Pretraining}
BERT (Bidirectional Encoder Representations from
Transformers) \cite{devlin-etal-2019-bert}, was designed to pretrain deep bidirectional representations of tokens from unlabeled text by jointly conditioning on both left and right context in all layers. This was achieved by using a "masked language model" (MLM), whose pretraining objective is to predict the randomly masked tokens of the sequence input. As a result, the pretrained BERT model can be finetuned with just one additional output layer to bring substantial performance gains for a wide range of language inference tasks, including sentiment classification, without extensive task-specific architecture modifications. Recently, Liu et al.~\cite{liu2019roberta} provided a replication study of BERT pretraining, and discovered that BERT was significantly undertrained, yet it can still match the performance of every model published following its inception. Thus, they presented additional insights on the design choices and training strategies of BERT and introduced alternative BERT-based models (RoBERTa) that record state-of-the-art results on similar tasks. They attributed their success to the use of a larger dataset for pretraining, and better design choices for MLM. They reported 0.948 F1 score of their RoBERTa\textsubscript{BASE} model for SST-2, a Stanford Sentiment Treebank (SST) dataset with binary labels (positive and negative) for sentiment analysis task. In comparison, BERT\textsubscript{BASE} "only" achieved 0.928 F1 score. Therefore, we choose RoBERTa as the pretraining procedure for our Twitter sentiment analysis model, as well as comparing its performance with BERT.

\subsubsection{Training and Evaluation}

We utilize the \textit{transformers}\footnote{https://git.io/JfUEh} library by \textit{huggingface} \cite{Wolf2019HuggingFacesTS}, which includes RoBERTa\textsubscript{BASE} and BERT\textsubscript{BASE} in Pytorch \cite{paszke2019pytorch}, and implement the sentiment analysis models with an additional linear layer on top of the pretrained model's outputs. The AdamW optimizer \cite{loshchilov2017decoupled} is used to optimize the cross-entropy loss function. We also use the \textit{fastai}\footnote{https://www.fast.ai/} API \cite{howard2020fastai}'s deep learning wrapper for Pytorch, which allows us to split the model's layers into groups, in order to use discriminative finetuning and slanted triangular learning rates \cite{howard-ruder-2018-universal} for task-specific features (i.e. learning word embeddings, learning context embeddings, and learning sentiment outputs).

We train and evaluate our models on the SemEval-2017 Task 4A dataset for Twitter message sentiment classification on a 3-point scale: Negative, Neutral, and Positive (Table \ref{tab2}). In the end, both of our classifier models substantially outperform the top two performers of SemEval-2017 (Table \ref{tab3}) on the test dataset. In particular, the model with RoBERTa pretraining achieves above 0.8 Macro-F1 score, which demonstrates its robustness for Twitter sentiment classification, and applicability for exploring the sentiments of our COVID-19 tweets. Since we implement the RoBERTa model to classify ternary sentiment labels, the drop in performance compared to Liu et al. ~\cite{liu2019roberta} is within expectation. The confusion matrix for our RoBERTa model is given in Table \ref{tab4}.

\begin{table}[htbp]
\caption{SemEval-2017 Task 4A Dataset Statistics}
\begin{center}
\begin{tabular}{c c c c c}
\hline
Dataset & \textbf{Negative}  & \textbf{Neutral} & \textbf{Positive} & Total \\
 \hline
Train & 7,838(15.6\%) & 22,586(44.9\%) & 19,896(39.5\%) & 50,320\\
Test & 3,959(32.4\%) & 5,894(48.2\%) & 2,365(19,4\%) & 12,218\\
\hline
\end{tabular}
\label{tab2}
\end{center}
\vspace{-0.2cm}
\end{table}

\begin{table}[htbp]
\caption{Performance Comparison between Previous Methods on Twitter Sentiment Classification and Ours}
\begin{center}
\begin{tabular}{c c c c}
\hline
Model & Accuracy  & Macro-F1 Score \\
 \hline
RoBERTa & \textbf{0.806} &  \textbf{0.806}\\
BERT & 0.757 & 0.757 \\
LSTMs$+$CNNs\cite{cliche-2017-bb} & $-$ & 0.685 \\
BiLSTMs$+$Attention\cite{baziotis-etal-2017-datastories-semeval} & $-$ & 0.677 \\
\hline
\end{tabular}
\label{tab3}
\end{center}
\vspace{-0.2cm}
\end{table}

\begin{table}[htbp]
\caption{Confusion Matrix of RoBERTa Model}
\begin{center}
\begin{tabular}{c c c c}
\hline
 & negative & neutral  & positive \\
 \hline
negative & 3315 & 843 & 23\\
neutral & 622 & 4643 & 458 \\
positive & 22 & 408 & 1884 \\
\hline
\end{tabular}
\label{tab4}
\end{center}
\vspace{-0.2cm}
\end{table}


\subsection{Analysis of Results}
\subsubsection{A depressing outlook of COVID-19}
\begin{figure}[htbp]
\centerline{\includegraphics[scale = 0.42]{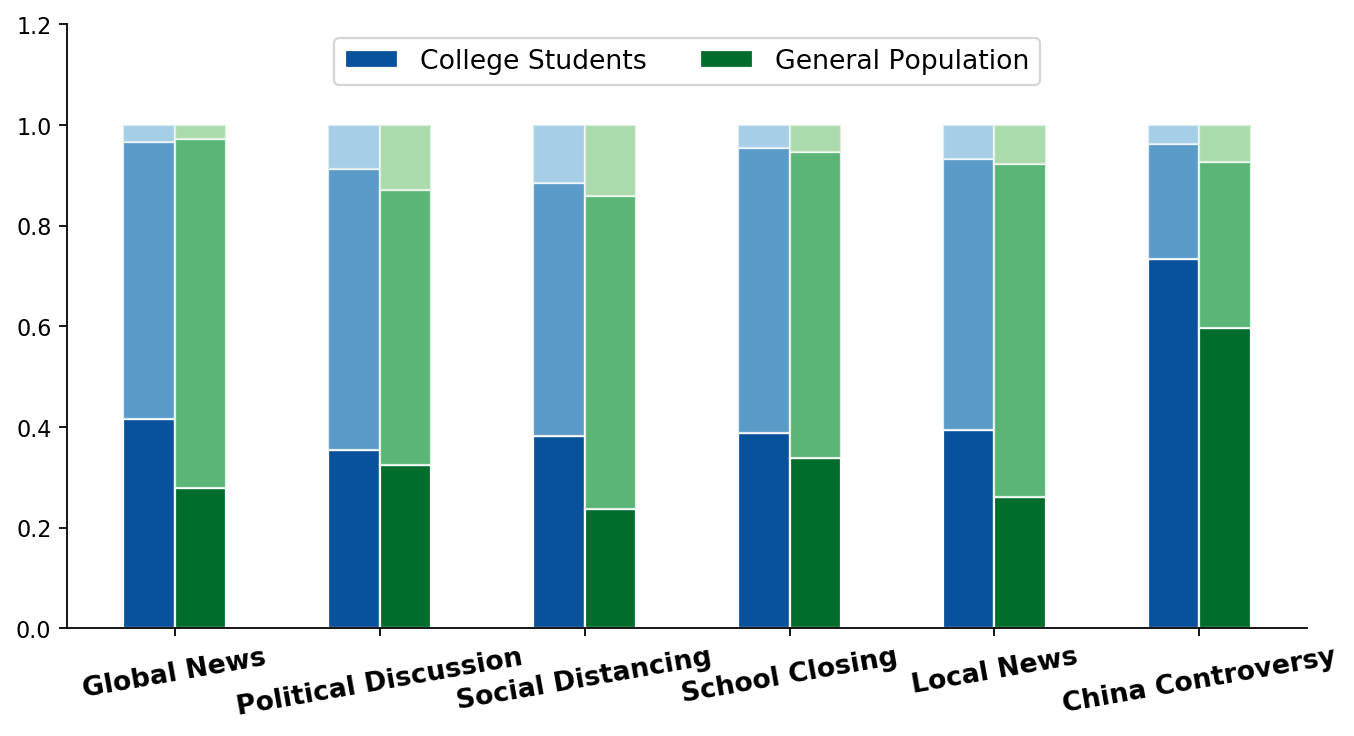}}
\caption{Sentiment Distributions (\%) towards the 6 Most Frequent COVID-19 Topics.  Percentage Blocks from Bottom to Top: Negative, Neutral, Positive.}
\label{fig6}
\end{figure}

Overall, a very small percentage of positive sentiments are expressed among the COVID-19 tweets (lightest-colored blocks of Figure \ref{fig6}).  In addition, more than one in five people of our user samples discussed COVID-19 related issues in a negative light. Considering that 2,281 out of a million of the U.S population are physically affected by COVID-19, which is already dangerous, the amount of negativity exhibited on Twitter is very alarming as well. Evidently, not only is the COVID-19 pandemic a health and safety hazard, it also has gloom-ridden impacts on our society. Moreover, for the topic related to the "Chinese virus" controversy, there is an overwhelming number of negative responses. We can see from Figure \ref{fig4} that "racist" is the 3rd most frequent keywords of this topic, which suggests that many of Twitter users associated calling "coronavirus" the "Chinese virus" with racism.

\subsubsection{College students respond more negatively to COVID-19}
An important trend that outweighs the rest of our results is that there is a significantly higher percentage among the student population expressing negative sentiments towards the central issues of COVID-19, especially on news related to the spread of the pandemic and social distancing. This shows that our analysis of the data is consistent with the our speculation on the impacts of the COVID-19 crisis on our younger population. College students are likely to express negative feelings towards how social distancing and school closure are affecting their work and study environments. Moreover, they tend to be subject to more negative emotions upon receiving news of the outbreak, which might be due to the subsequent implications of these issues on those more related to their lives.
\subsubsection{Negativity among College Students through the Topic Modeling Microscope}
We focus on examining the subtopics of School Closing, which is of high concern among college students, and on Social Distancing and China Controversy, where highest gaps in the negative sentiments between college students and the general population are observed (14.5\% and 13.8\% absolute difference in percentage of negative tweets respectively). In addition, only negative and positive tweets are considered because they provide the most meaningful contexts associated with their sentiments. In general, non-neutral tweets on the Social Distancing and School Closing topics express worrying emotions towards COVID-19, and all the tweets revealing concerns on school closure are negative. Moreover, many students exhibited aggression to the foreign community, blaming them for the current disruptions in their lives as a result of social distancing. College students also disclosed details of their online learning experience, and mostly showed dislikes for remote learning (81.3\%). To reflect on the responses of college students on COVID-19 in a more positive light, it is encouraging that our college community remains aware and vocal on the racism problem related to the "Chinese virus" controversy, which sends a powerful message on the public's intolerance of racist behaviors on social media for the betterment of our society.

These findings shed new lights on an emerging direction of racism problems in the U.S. during the COVID-19 outbreak, especially related to the East Asian community, in addition to the existing discussions in the literature that focus primarily on discrimination towards African American, Asian American, and more recently, Muslim population \cite{swim2003african,museus2015continuing,Guhin2018}. Also, in addition to addressing the uneasy feelings and barriers restricting the learning experience among students during the crisis, the prevention of racism-charged hate speeches is an important task for educational institutions to protect their students.

\begin{table}[htbp]
\caption{Subtopics of Social Distancing}
\begin{center}
\begin{tabular}{|p{1.4cm}|p{4.8cm}|p{1cm}|}
\hline
Subtopic Label & Subtopic Keywords  & Negative Tweets\\
\hline
Showing aggression & asia, people, stay, home, worse\_european, everyone, piss, fucking, work, fight & 81.5\% \\
\hline
Detailing precautions & worry, safe, world, tour, stay, people, cancel, take, wash\_hand, precaution &  65.1\%\\
\hline
Expressing concerns & sick, know, go, work, see, watch, really, think, grocery\_store, family & 85.5\% \\
\hline
\end{tabular}
\label{tab5}
\end{center}
\vspace{-0.2cm}
\end{table}

\begin{table}[htbp]
\caption{Subtopics of School Closing}
\begin{center}
\begin{tabular}{|p{1.4cm}|p{5cm}|p{1cm}|}
\hline
Subtopic Label & Subtopic Keywords  & Negative Tweets\\
\hline
Detailing current situations & knock, follow\_government\_instruction, feel, survival\_rate\_whole, country\_panicking, get, people, fuck, right, week, campus &  98.5\%\\
\hline
Detailing remote study & school, close, shut, student, find, get, campus, live\_streaming\_instead, email, anymore &  81.3\%\\
\hline
Expressing concerns & people, due, cancel, go, concern, tell, week, imagine, nasty, break &  100\%\\
\hline
\end{tabular}
\label{tab6}
\end{center}
\vspace{-0.2cm}
\end{table}

\begin{table}[htbp]
\caption{Subtopics of China Controversy}
\begin{center}
\begin{tabular}{|p{1.4cm}|p{5cm}|p{1cm}|}
\hline
Subtopic Label & Subtopic Keywords  & Negative Tweets\\
\hline
Calling out racism & chinese, president, call, racist, refer, flu, reply, people, fuck, trump &  95.1\% \\
\hline
Addressing attitudes & people, take, perspective, sick, seriously, asian, friend, know, ass, time & 97.3\% \\
\hline
Detailing public response & call, guy, get, keep, chinese, think, remember, wuhan, response, piss & 91.7\% \\
\hline
\end{tabular}
\label{tab7}
\end{center}
\vspace{-0.2cm}
\end{table}

\section{Conclusions and Future Work}
We have analyzed 73,787 tweets from 12,776 Twitter college followers who posted tweets related the COVID-19 pandemic, in terms of the outstanding topics on several social issues. We find significant differences in the sentiments expressed towards those topics between the users who are identified as colleges students and those of the general population. College students tend to focus their discussions on topics closely surrounding their living environment, such as school closure and local news. Although the percentages of positive COVID-19 tweets are very low for both demographics, college students are shown to be significantly more negative. In addition, microscopic examination of the positive and negative tweets reveals their overwhelmingly troubled feelings amidst the spread of COVID-19, as well as unfavorable reactions to the disruption in their lives such as racism-charged aggression. Moreover, we discover a shift in the target of racism during COVID-19 towards the East Asian community, which the majority of college students and the general public are against.

Since high accuracy is achieved in both of our demographic and sentiment classification models, future studies may collect larger datasets to achieve better performance. In addition, this research mainly focuses on high-level attributes of tweets such as topic models and sentiments in understanding the characteristics of users who discussed social issues associated with COVID-19. Analysis on more fine-grained linguistic information, such as emotion, hate speech, and racism detection can be performed to gain further insights on the more specific COVID-19 related issues detailed in our study.

\bibliographystyle{IEEEtran}
\bibliography{IEEEexample}

\begin{thebibliography}{10}
\providecommand{\url}[1]{#1}
\csname url@samestyle\endcsname
\providecommand{\newblock}{\relax}
\providecommand{\bibinfo}[2]{#2}
\providecommand{\BIBentrySTDinterwordspacing}{\spaceskip=0pt\relax}
\providecommand{\BIBentryALTinterwordstretchfactor}{4}
\providecommand{\BIBentryALTinterwordspacing}{\spaceskip=\fontdimen2\font plus
\BIBentryALTinterwordstretchfactor\fontdimen3\font minus
  \fontdimen4\font\relax}
\providecommand{\BIBforeignlanguage}[2]{{%
\expandafter\ifx\csname l@#1\endcsname\relax
\typeout{** WARNING: IEEEtran.bst: No hyphenation pattern has been}%
\typeout{** loaded for the language `#1'. Using the pattern for}%
\typeout{** the default language instead.}%
\else
\language=\csname l@#1\endcsname
\fi
#2}}
\providecommand{\BIBdecl}{\relax}
\BIBdecl

\bibitem{fedynich2013teaching}
L.~V. Fedynich, ``Teaching beyond the classroom walls: The pros and cons of
  cyber learning.'' \emph{Journal of Instructional Pedagogies}, vol.~13, 2013.

\bibitem{morrison2019designing}
G.~R. Morrison, S.~J. Ross, J.~R. Morrison, and H.~K. Kalman, \emph{Designing
  effective instruction}.\hskip 1em plus 0.5em minus 0.4em\relax John Wiley \&
  Sons, 2019.

\bibitem{sahu2020closure}
P.~Sahu, ``Closure of universities due to coronavirus disease 2019 (covid-19):
  Impact on education and mental health of students and academic staff,''
  \emph{Cureus}, vol.~12, no.~4, 2020.

\bibitem{rao2010classifying}
D.~Rao, D.~Yarowsky, A.~Shreevats, and M.~Gupta, ``Classifying latent user
  attributes in twitter,'' in \emph{Proceedings of the 2nd international
  workshop on Search and mining user-generated contents}, 2010, pp. 37--44.

\bibitem{bergsma2013using}
S.~Bergsma and B.~Van~Durme, ``Using conceptual class attributes to
  characterize social media users,'' in \emph{Proceedings of the 51st Annual
  Meeting of the Association for Computational Linguistics (Volume 1: Long
  Papers)}, 2013, pp. 710--720.

\bibitem{nguyen2013old}
D.~Nguyen, R.~Gravel, D.~Trieschnigg, and T.~Meder, ``" how old do you think i
  am?" a study of language and age in twitter,'' in \emph{Seventh International
  AAAI Conference on Weblogs and Social Media}, 2013.

\bibitem{sloan2015tweets}
L.~Sloan, J.~Morgan, P.~Burnap, and M.~Williams, ``Who tweets? deriving the
  demographic characteristics of age, occupation and social class from twitter
  user meta-data,'' \emph{PloS one}, vol.~10, no.~3, 2015.

\bibitem{beller2014ma}
C.~Beller, R.~Knowles, C.~Harman, S.~Bergsma, M.~Mitchell, and B.~Van~Durme,
  ``I’ma belieber: Social roles via self-identification and conceptual
  attributes,'' in \emph{Proceedings of the 52nd Annual Meeting of the
  Association for Computational Linguistics (Volume 2: Short Papers)}, 2014,
  pp. 181--186.

\bibitem{hanson2013tweaking}
C.~L. Hanson, S.~H. Burton, C.~Giraud-Carrier, J.~H. West, M.~D. Barnes, and
  B.~Hansen, ``Tweaking and tweeting: exploring twitter for nonmedical use of a
  psychostimulant drug (adderall) among college students,'' \emph{Journal of
  medical Internet research}, vol.~15, no.~4, p. e62, 2013.

\bibitem{he2016using}
L.~He, L.~Murphy, and J.~Luo, ``Using social media to promote stem education:
  Matching college students with role models,'' in \emph{Joint European
  Conference on Machine Learning and Knowledge Discovery in Databases}.\hskip
  1em plus 0.5em minus 0.4em\relax Springer, 2016, pp. 79--95.

\bibitem{wang2019demographic}
Z.~Wang, S.~A. Hale, D.~Adelani, P.~A. Grabowicz, T.~Hartmann, F.~Fl{\"o"}ck,
  and D.~Jurgens, ``Demographic inference and representative population
  estimates from multilingual social media data,'' in \emph{Proceedings of the
  2019 World Wide Web Conference}.\hskip 1em plus 0.5em minus 0.4em\relax ACM,
  2019.

\bibitem{giachanou2016like}
A.~Giachanou and F.~Crestani, ``Like it or not: A survey of twitter sentiment
  analysis methods,'' \emph{ACM Computing Surveys (CSUR)}, vol.~49, no.~2, pp.
  1--41, 2016.

\bibitem{korenek2014sentiment}
P.~Korenek and M.~{\v{S}}imko, ``Sentiment analysis on microblog utilizing
  appraisal theory,'' \emph{World Wide Web}, vol.~17, no.~4, pp. 847--867,
  2014.

\bibitem{tang2014learning}
D.~Tang, F.~Wei, N.~Yang, M.~Zhou, T.~Liu, and B.~Qin, ``Learning
  sentiment-specific word embedding for twitter sentiment classification,'' in
  \emph{Proceedings of the 52nd Annual Meeting of the Association for
  Computational Linguistics (Volume 1: Long Papers)}, 2014, pp. 1555--1565.

\bibitem{pennington2014glove}
J.~Pennington, R.~Socher, and C.~D. Manning, ``Glove: Global vectors for word
  representation,'' in \emph{Proceedings of the 2014 conference on empirical
  methods in natural language processing (EMNLP)}, 2014, pp. 1532--1543.

\bibitem{cliche-2017-bb}
\BIBentryALTinterwordspacing
M.~Cliche, ``{BB}{\_}twtr at {S}em{E}val-2017 task 4: Twitter sentiment
  analysis with {CNN}s and {LSTM}s,'' in \emph{Proceedings of the 11th
  International Workshop on Semantic Evaluation ({S}em{E}val-2017)}.\hskip 1em
  plus 0.5em minus 0.4em\relax Vancouver, Canada: Association for Computational
  Linguistics, Aug. 2017, pp. 573--580. [Online]. Available:
  \url{https://www.aclweb.org/anthology/S17-2094}
\BIBentrySTDinterwordspacing

\bibitem{baziotis-etal-2017-datastories-semeval}
\BIBentryALTinterwordspacing
C.~Baziotis, N.~Pelekis, and C.~Doulkeridis, ``{D}ata{S}tories at
  {S}em{E}val-2017 task 4: Deep {LSTM} with attention for message-level and
  topic-based sentiment analysis,'' in \emph{Proceedings of the 11th
  International Workshop on Semantic Evaluation ({S}em{E}val-2017)}.\hskip 1em
  plus 0.5em minus 0.4em\relax Vancouver, Canada: Association for Computational
  Linguistics, Aug. 2017, pp. 747--754. [Online]. Available:
  \url{https://www.aclweb.org/anthology/S17-2126}
\BIBentrySTDinterwordspacing

\bibitem{rosenthal2019semeval}
S.~Rosenthal, N.~Farra, and P.~Nakov, ``Semeval-2017 task 4: Sentiment analysis
  in twitter,'' \emph{arXiv preprint arXiv:1912.00741}, 2019.

\bibitem{devlin-etal-2019-bert}
\BIBentryALTinterwordspacing
J.~Devlin, M.-W. Chang, K.~Lee, and K.~Toutanova, ``{BERT}: Pre-training of
  deep bidirectional transformers for language understanding,'' in
  \emph{Proceedings of the 2019 Conference of the North {A}merican Chapter of
  the Association for Computational Linguistics: Human Language Technologies,
  Volume 1 (Long and Short Papers)}.\hskip 1em plus 0.5em minus 0.4em\relax
  Minneapolis, Minnesota: Association for Computational Linguistics, Jun. 2019,
  pp. 4171--4186. [Online]. Available:
  \url{https://www.aclweb.org/anthology/N19-1423}
\BIBentrySTDinterwordspacing

\bibitem{liu2019roberta}
Y.~Liu, M.~Ott, N.~Goyal, J.~Du, M.~Joshi, D.~Chen, O.~Levy, M.~Lewis,
  L.~Zettlemoyer, and V.~Stoyanov, ``Roberta: A robustly optimized bert
  pretraining approach,'' \emph{arXiv preprint arXiv:1907.11692}, 2019.

\bibitem{esuli2006sentiwordnet}
A.~Esuli and F.~Sebastiani, ``Sentiwordnet: A publicly available lexical
  resource for opinion mining.'' in \emph{LREC}, vol.~6.\hskip 1em plus 0.5em
  minus 0.4em\relax Citeseer, 2006, pp. 417--422.

\bibitem{chen2020covid}
E.~Chen, K.~Lerman, and E.~Ferrara, ``Covid-19: The first public coronavirus
  twitter dataset,'' \emph{arXiv preprint arXiv:2003.07372}, 2020.

\bibitem{manning-etal-2014-stanford}
\BIBentryALTinterwordspacing
C.~Manning, M.~Surdeanu, J.~Bauer, J.~Finkel, S.~Bethard, and D.~McClosky,
  ``The {S}tanford {C}ore{NLP} natural language processing toolkit,'' in
  \emph{Proceedings of 52nd Annual Meeting of the Association for Computational
  Linguistics: System Demonstrations}.\hskip 1em plus 0.5em minus 0.4em\relax
  Baltimore, Maryland: Association for Computational Linguistics, Jun. 2014,
  pp. 55--60. [Online]. Available:
  \url{https://www.aclweb.org/anthology/P14-5010}
\BIBentrySTDinterwordspacing

\bibitem{sebastiani2002machine}
F.~Sebastiani, ``Machine learning in automated text categorization,'' \emph{ACM
  computing surveys (CSUR)}, vol.~34, no.~1, pp. 1--47, 2002.

\bibitem{church1990word}
K.~W. Church and P.~Hanks, ``Word association norms, mutual information, and
  lexicography,'' \emph{Computational linguistics}, vol.~16, no.~1, pp. 22--29,
  1990.

\bibitem{NIPS2010_3902}
\BIBentryALTinterwordspacing
M.~Hoffman, F.~R. Bach, and D.~M. Blei, ``Online learning for latent dirichlet
  allocation,'' in \emph{Advances in Neural Information Processing Systems 23},
  J.~D. Lafferty, C.~K.~I. Williams, J.~Shawe-Taylor, R.~S. Zemel, and
  A.~Culotta, Eds.\hskip 1em plus 0.5em minus 0.4em\relax Curran Associates,
  Inc., 2010, pp. 856--864. [Online]. Available:
  \url{http://papers.nips.cc/paper/3902-online-learning-for-latent-dirichlet-allocation.pdf}
\BIBentrySTDinterwordspacing

\bibitem{maaten2008visualizing}
L.~v.~d. Maaten and G.~Hinton, ``Visualizing data using t-sne,'' \emph{Journal
  of machine learning research}, vol.~9, no. Nov, pp. 2579--2605, 2008.

\bibitem{Wolf2019HuggingFacesTS}
T.~Wolf, L.~Debut, V.~Sanh, J.~Chaumond, C.~Delangue, A.~Moi, P.~Cistac,
  T.~Rault, R.~Louf, M.~Funtowicz, and J.~Brew, ``Huggingface's transformers:
  State-of-the-art natural language processing,'' \emph{ArXiv}, vol.
  abs/1910.03771, 2019.

\bibitem{paszke2019pytorch}
A.~Paszke, S.~Gross, F.~Massa, A.~Lerer, J.~Bradbury, G.~Chanan, T.~Killeen,
  Z.~Lin, N.~Gimelshein, L.~Antiga \emph{et~al.}, ``Pytorch: An imperative
  style, high-performance deep learning library,'' in \emph{Advances in Neural
  Information Processing Systems}, 2019, pp. 8024--8035.

\bibitem{loshchilov2017decoupled}
I.~Loshchilov and F.~Hutter, ``Decoupled weight decay regularization,''
  \emph{arXiv preprint arXiv:1711.05101}, 2017.

\bibitem{howard2020fastai}
J.~Howard and S.~Gugger, ``Fastai: A layered api for deep learning,''
  \emph{Information}, vol.~11, no.~2, p. 108, 2020.

\bibitem{howard-ruder-2018-universal}
\BIBentryALTinterwordspacing
J.~Howard and S.~Ruder, ``Universal language model fine-tuning for text
  classification,'' in \emph{Proceedings of the 56th Annual Meeting of the
  Association for Computational Linguistics (Volume 1: Long Papers)}.\hskip 1em
  plus 0.5em minus 0.4em\relax Melbourne, Australia: Association for
  Computational Linguistics, Jul. 2018, pp. 328--339. [Online]. Available:
  \url{https://www.aclweb.org/anthology/P18-1031}
\BIBentrySTDinterwordspacing

\bibitem{swim2003african}
J.~K. Swim, L.~L. Hyers, L.~L. Cohen, D.~C. Fitzgerald, and W.~H. Bylsma,
  ``African american college students’ experiences with everyday racism:
  Characteristics of and responses to these incidents,'' \emph{Journal of Black
  psychology}, vol.~29, no.~1, pp. 38--67, 2003.

\bibitem{museus2015continuing}
S.~D. Museus and J.~J. Park, ``The continuing significance of racism in the
  lives of asian american college students,'' \emph{Journal of College Student
  Development}, vol.~56, no.~6, pp. 551--569, 2015.

\bibitem{Guhin2018}
J.~Guhin, ``Colorblind islam: the racial hinges of immigrant muslims in the
  united states,'' \emph{Social Inclusion}, vol.~6, no.~2, pp. 87--97, 2018.

\end{thebibliography}

\end{document}